\documentclass[british,english]{article}
\usepackage[a4paper]{geometry}
\usepackage{color}
\usepackage{babel}
\usepackage{array}
\usepackage{float}
\usepackage{mathtools}
\usepackage{multicol}
\usepackage{amsmath}
\usepackage{amssymb}
\usepackage{graphicx}
\usepackage[numbers,sort]{natbib}
\usepackage[unicode=true,bookmarks=true,bookmarksnumbered=false,bookmarksopen=false,breaklinks=false,pdfborder={0 0 0},pdfborderstyle={},backref=false,colorlinks=true]{hyperref}
\usepackage{paracol}
\usepackage{enumitem}
\makeatletter
\hypersetup{
urlcolor = blue,
linkcolor = blue,
citecolor = blue
}
\makeatother

\begin{document}

\title{\noindent DistributedFBA.jl: High-level, high-performance flux balance analysis in Julia}

\author{Laurent Heirendt, Ronan M.T. Fleming, Ines Thiele}
\date{\vspace{-5ex}}
\maketitle

\vspace*{1ex}
\paragraph*{Motivation}

Flux balance analysis, and its variants, are widely used methods for
predicting steady-state reaction rates in biochemical reaction networks.
The exploration of high dimensional networks with such methods is
currently hampered by software performance limitations.

\paragraph*{Results}

\emph{DistributedFBA.jl} is a high-level, high-performance, open-source
implementation of flux balance analysis in Julia. It is tailored to
solve multiple flux balance analyses on a subset or all the reactions
of large and huge-scale networks, on any number of threads or nodes.

\paragraph*{Availability}

The code and benchmark data are freely available on \emph{\href{http://github.com/opencobra/COBRA.jl}{github.com/opencobra/COBRA.jl}.}
The documentation can be found at \emph{\href{http://opencobra.github.io/COBRA.jl}{opencobra.github.io/COBRA.jl}.}

\paragraph*{Contact}

\noindent \emph{\href{http://ines.thiele@uni.lu}{ines.thiele@uni.lu}}

\noindent 
\begin{multicols}{2}

\section{Introduction}

Constraint-Based Reconstruction and Analysis (COBRA)~\citep{Pal15}
is a widely used approach for modeling genome-scale biochemical networks
and integrative analysis of omics data in a network context. All COBRA
predictions are derived from optimisation problems, typically formulated
in the form
\begin{equation}
\begin{array}{ll}
\underset{v\in\mathbb{R}^{n}}{\textrm{min}} & \psi(v)\\
\;\text{s.t.} & Sv=b\\
 & Cv\le d\\
 & l\leq v\leq u,
\end{array}\label{eq1}
\end{equation}
\noindent where $v\in\mathbb{R}^{n}$ represents the rate of each
biochemical reaction, $\psi:\mathbb{R}^{n}\rightarrow\mathbb{R}$
is a lower semi-continuous and convex function, $S\in\mathbb{R}^{m\times n}$
is a stoichiometric matrix for $m$ molecular species and $n$ reactions,
and $b$ is a vector of known metabolic exchanges. Additional linear
inequalities (expressed as a system of equations with matrix $C$
and vector $d$) may be used to constrain combinations of reaction
rates and keep reactions between upper and lower bounds, $u$ and
$l$ respectively.

In flux balance analysis (FBA) one obtains a steady-state
by choosing a coefficient vector $c\in\mathbb{R}^{n}$ and letting
$\psi(v) := c^{T}v$ and $b := 0$. However, the biologically
correct coefficient vector is usually not known, so exploration of
the set of steady states relies on the embarrassingly parallel problem
of solving (\ref{eq1}) for many $c$. Moreover, while $c^{T}v^{\star}$
is unique for an optimal flux vector $v^{\star}$, there may
be alternate optimal solutions. In flux variability analysis (FVA),
one finds the extremes for each optimal reaction rate by choosing
a coefficient vector $d\in\mathbb{R}^{n}$ with one nonzero entry,
then minimising and maximising $\psi(v) := d^{T}v$, subject to the 
additional constraint $d^{T}v\geq\gamma\cdot c^{T}v^{\star}$
for each reaction in turn ($\gamma\in]0,1[$).

For kilo-scale models ($n\simeq1,000$), the $2n$
linear optimisation problems required for FVA can currently be solved
efficiently using existing methods, e.g., \emph{FVA} of the COBRA Toolbox, 
\emph{fastFVA}, or the \emph{COBRApy} implementation~\citep{Gud10,Ebr13,Sch11}.
However, these implementations perform best when using only one 
computing node with a few cores, which becomes a temporal limiting 
factor when exploring the steady state solution space of larger models. 
Julia is a high-level, high-performance dynamic programming language
for technical computing~\citep{Bez14}. Here, we exploit 
Julia to distribute sets of FBA problems and compare its performance
to existing implementations.

\section{Overview and implementation}

\emph{DistributedFBA.jl}, part of a novel \emph{COBRA.jl} package,
is implemented in Julia, and makes use of the high-level interface
\emph{MathProgBase.jl}~\citep{Lub15} (see Supplementary
Material). A key feature is the integrated capability of spawning
synchronously any number of processes to local and remote workers.
\emph{COBRA.jl} extends the COBRA Toolbox~\citep{Sch11}
while existing COBRA models~\citep{Ort10} can be input.

\section{Benchmark results}

\emph{DistributedFBA.jl} and \emph{fastFVA}~\citep{Gud10} 
were benchmarked on a set of models of varying dimension
(Table~\ref{tab:Tab1}). All experiments were run
on several DELL R630 computing nodes with 2x36 threads and 768GB
RAM running Linux. As Julia is a just-in-time language, pre-compilation
(warm-up) was done on a small-scale model before benchmarking~\citep{Ort10}.
The creation of a parallel pool of workers and the time to spawn
the processes are not considered in the reported times.

\begin{table}[H]
\noindent \begin{centering}
{\footnotesize{}}%
\begin{tabular}{>{\raggedright}p{2.5cm}>{\raggedright}p{1.5cm}>{\raggedright}p{1.5cm}l}
\emph{Model} & $m$ & $n$ & \emph{Ref.}\tabularnewline
\hline 
Recon1  			& $2,785$ 	& $3,820$ 	& \citep{Ree06}\tabularnewline
Recon2  			& $5,063$ 	& $7,440$ 	& \citep{Nor16}\tabularnewline
Recon3  			& $7,866$ 	& $12,566$ 	& \citep{Bru16}\tabularnewline
Recon2+11M  	& $19,714$ 	& $28,199$ 	& \citep{Thi16}\tabularnewline
Multi-organ    	& $47,123$ 	& $61,230$ 	& \citep{Thi16-1}\tabularnewline
\hline 
\end{tabular}
\par\end{centering}
{\caption{Sizes of $S$ for benchmark models.}\label{tab:Tab1}}
\end{table}
The serial performance of both implementations is within 10\%. 
The uninodal performance of \emph{fastFVA} is slightly higher on a 
few threads, but the performance of \emph{distributedFBA.jl} is superior
for a higher number of threads on a single node (Fig.~\ref{fig:Fig1}~A). 
The way the FBA problems are distributed among workers (distribution 
strategy $s$, see Supplementary Material) yields an additional speedup of 
10-20\% on a larger number of threads.
\begin{figure}[H]
\centerline{\includegraphics[width=\columnwidth]{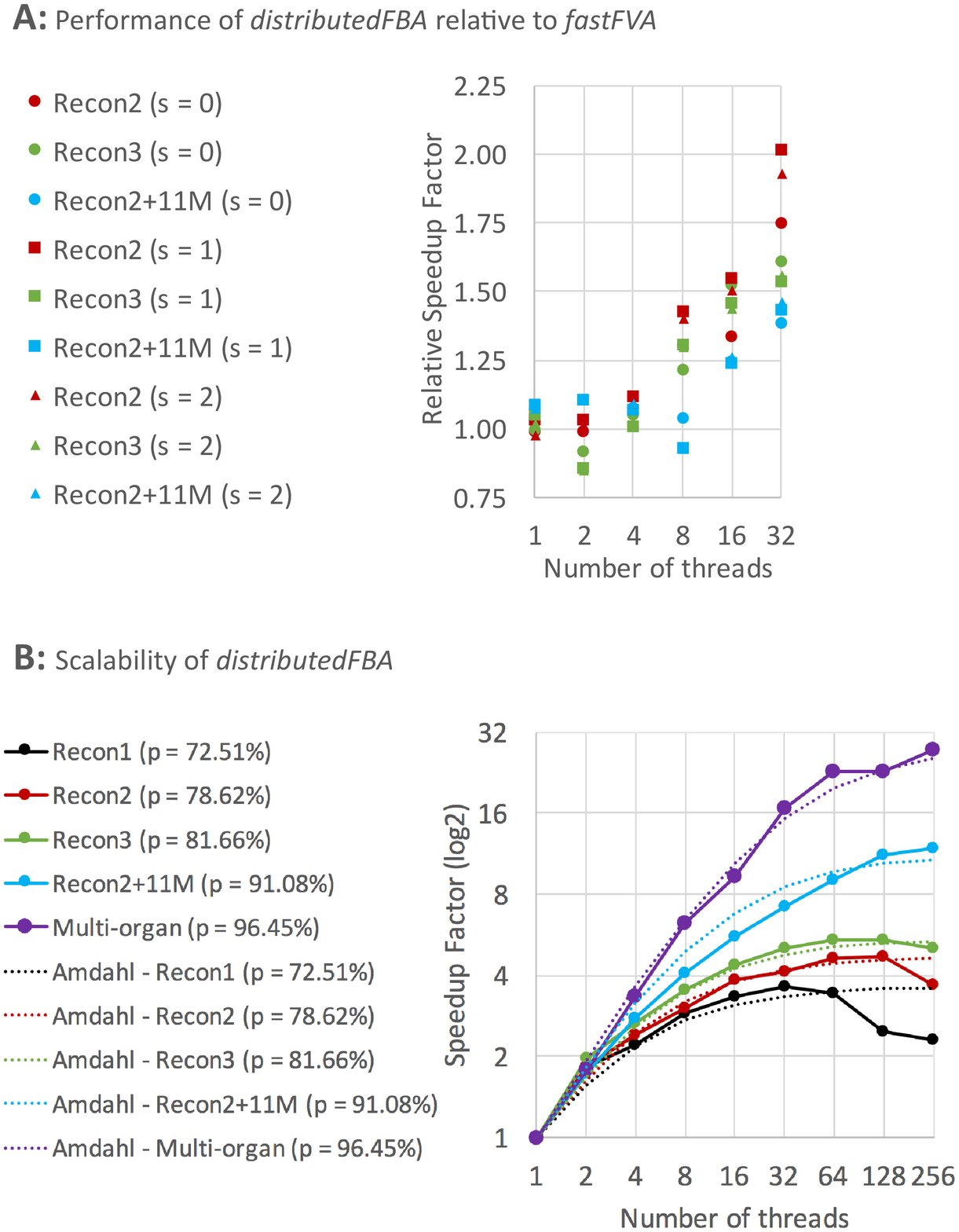}}
\caption{Performance of \emph{distributedFBA} for the benchmark models 
given in Table ~\ref{tab:Tab1}. {\bf A}: Speedup factor relative to
\emph{fastFVA} as a function of threads and distribution strategy $s$ (1 node).
{\bf B}: Multi-nodal speedup in latency and Amdahl's law ($s=0$).
\label{fig:Fig1}}
\end{figure}
According to Amdahl's law, the theoretical speedup
factor is $\left(1-p+\frac{p}{N}\right)^{-1}$, where $N$ is the
number of cores and $p$ is the fraction  of the code (including
the model) that can be parallelised. The fraction $p$ increases
with an increasing model size (Fig.~\ref{fig:Fig1}~B).
The maximum speedup factor for a very large number of cores $N$ is
$\left(1-p\right)^{-1}$. This demonstrates that for high dimensional
models, it is critical to have a large number of threads on multiple
high-memory nodes to accrue a significant speedup.

\section{Discussion}

The multi-nodal performance of \emph{distributedFBA.jl} is unparalleled:
the scalability of \emph{distributedFBA.jl} matches theoretical predictions,
and resources are optimally used. Key advantages are that the present
implementation is open-source, platform independent, and that no pool size
limits, memory, or node/thread limitations exist. Its uninodal performance is
similar to the performance of \emph{fastFVA} on a few threads,
and about 2-3 times higher on a larger number of threads. A key reason
is the direct parallelisation capabilities of Julia and the wrapper-free
interface to the solver. The unilingual and easy-to-use implementation
relies on solvers written in other languages, allows the analysis
of large and huge-scale biochemical networks in a timely manner, and
lifts the analysis possibilities in the COBRA community to another level.

\section*{Acknowledgement}

\noindent\emph{Funding:} This study was funded the National Centre
of Excellence in Research (NCER) on Parkinson's disease
and by the U.S. Department of Energy, Offices of Advanced Scientific
Computing Research and the Biological and Environmental Research as
part of the Scientific Discovery Through Advanced Computing program,
grant \#DE-SC0010429.
\vspace{0.1cm}

\noindent\emph{Conflict of Interest:} none declared.
\setlength{\bibsep}{0pt plus 0.25ex}
\def\bibindent{1mm}
\def\bibfont{\scriptsize}
\bibliographystyle{unsrt}
\begin{scriptsize}

\end{scriptsize}
\end{multicols}

\pagebreak{}

\section*{Supplementary Material}

\begin{figure}[H]
	\centering
	\includegraphics[width=\textwidth]{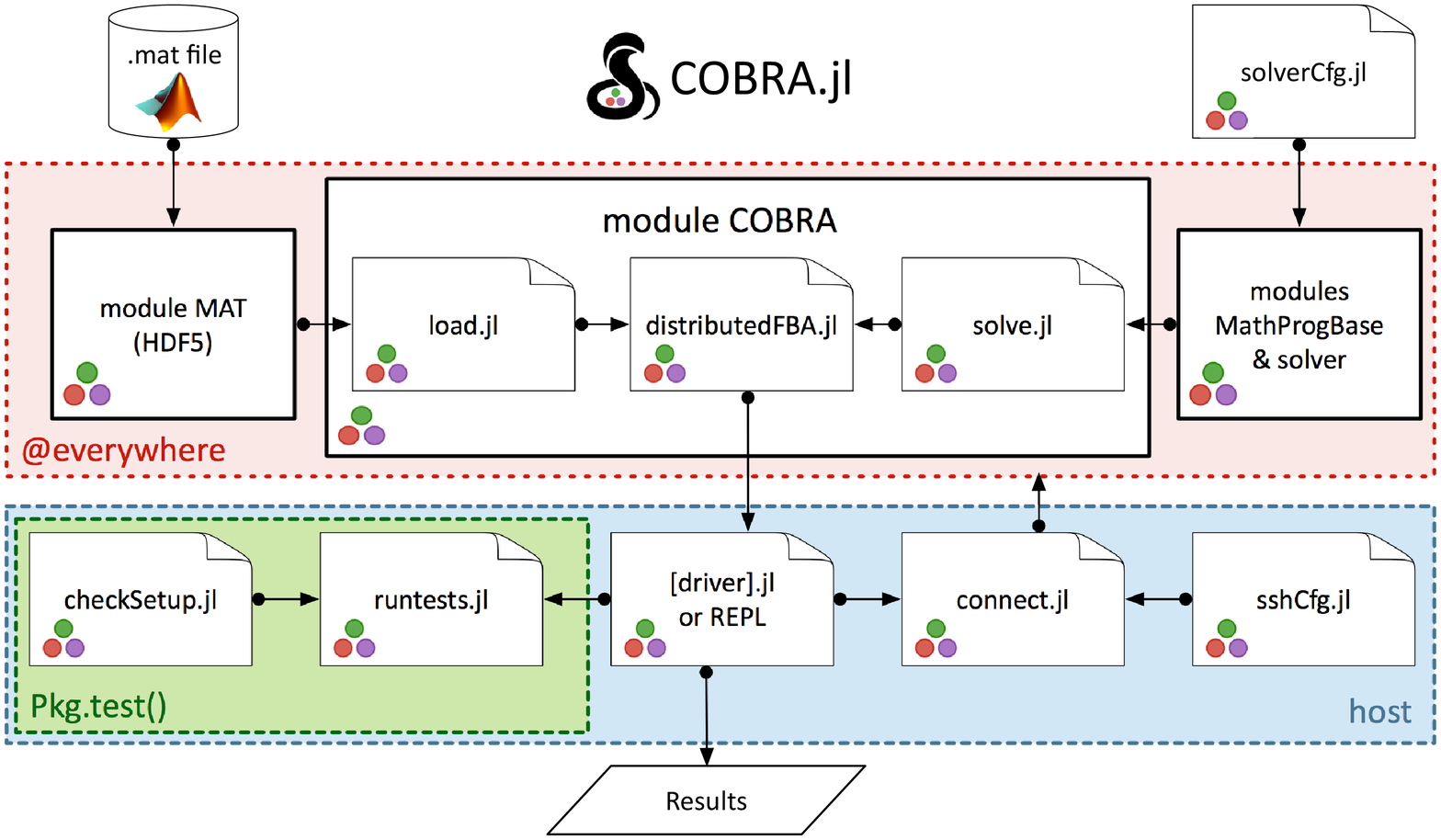}
	\caption{Overview of the \emph{COBRA.jl} package - v0.1.}\label{fig:OverviewPkg}
\end{figure}

\emph{DistributedFBA.jl} is part of \emph{COBRA.jl} (see Fig.~\ref{fig:OverviewPkg}).
The COBRA module wraps \emph{load.jl}, \emph{distributedFBA.jl}, and
\emph{solve.jl}. The input to the COBRA module is a \emph{.mat} file
that contains data of a COBRA model as defined in~\citep{Sch11}.
This HDF5 model is loaded using the \emph{MAT.jl} module~\citep{Kor12}.
Additionally, solver configuration parameters that are set in \emph{solverCfg.jl}.
are input to the COBRA module.
A parallel pool with either local or remote workers (using \emph{connect.jl})
may be created using either the Julia REPL or a driver. The COBRA
module and its dependencies, such as \emph{MathProgBase.jl}~\citep{Lub15}
and solver interfaces, are spawned from the host node to each worker
with the macro \emph{@everywhere}. This ensures that the full model
and the solver interfaces are available on each worker (including
the host), although only a subset of the FBA problems are solved on
each worker. The results are assembled on the host and fetched from
the workers independent of the size of the parallel pool.

The core functions for distributing and solving multiple FBA problems
are defined in the COBRA module. The main function within the COBRA
module is \emph{distributedFBA()} defined in \emph{distributedFBA.jl},
which loads the model from file (\emph{load.jl}: \emph{loadModel()}),
builds the LP model (\emph{solve.jl: buildCobraLP()}), and maximises
or minimises the LP problem (\emph{solve.jl: solveCobraLP()}) on the
spawned processes with a different set of FBA problems using \emph{distributedFBA.jl: loopFBA()}.
Before the LP problems are solved, additional constraints may be added
to the model using \emph{distributedFBA.jl: preFBA!()}. The FBA problems
are distributed using \emph{distributedFBA.jl: splitRange()} according
to the splitting strategy $s$, which is based on the sorted column
density vector $\rho_{c}$ of the stoichiometric matrix $S$:
\begin{itemize}
\item $s=0$ :\emph{Blind splitting}: default random distribution
\item $s=1$ : \emph{Extremal dense-and-sparse splitting}: every thread
receives dense and sparse reactions, starting from \emph{extremal}
indices of $\rho_{c}$
\item $s=2$ : \emph{Central dense-and-sparse splitting:} every thread receives
dense and sparse reactions, starting from the \emph{central} indices
of $\rho_{c}$
\end{itemize}
The COBRA module may be tested using \emph{runtests.jl}, which also
checks the computing node configuration (\emph{checkSetup.jl}), and
confirms that a compatible and working solver installation is present.
\end{document}